\documentclass[twocolumn,showpacs,preprintnumbers,amsmath,amssymb]{revtex4}
\date{March 2010}
\usepackage{epsfig}
\usepackage{latexsym}
\usepackage{amssymb}
\usepackage{booktabs}
\usepackage{graphicx}
\newcommand{\be}{\begin{equation}}
\newcommand{\ee}{\end{equation}}
\newcommand{\ba}{\begin{eqnarray}}
\newcommand{\ea}{\end{eqnarray}}
\newcommand{\bi}{\begin{itemize}}
\newcommand{\ei}{\end{itemize}}

\newcommand{\half}{{\textstyle\frac{1}{2}}}

\newcommand{\<}{\langle}
\renewcommand{\>}{\rangle}
\newcommand{\eq}{Eq.~}

\newcommand{\fig}{Fig.~}

\newcommand{\la}{\label}

\newcommand{\txts}{\textstyle}

\begin{document}
\preprint{MITP/13-032~~~~ HIM-2013-05}
\title{A new representation of the Adler function for lattice QCD}

\author{Anthony Francis, Benjamin J\"ager, Harvey~B.~Meyer, Hartmut Wittig}

\affiliation{\vspace{0.2cm}PRISMA Cluster of Excellence,
Institut f\"ur Kernphysik and Helmholtz~Institut~Mainz,
Johannes Gutenberg-Universit\"at Mainz,
D-55099 Mainz, Germany\vspace{0.2cm}}

\date{\today}

\begin{abstract}
We address several aspects of lattice QCD calculations of the hadronic vacuum polarization
and the associated Adler function. We implement a representation derived previously 
which allows one to access these phenomenologically important functions 
for a continuous set of virtualities, irrespective of the flavor structure of the current.
Secondly we present a theoretical analysis of the finite-size effects on our particular representation of 
the Adler function, based on the operator product expansion at large momenta 
and on the spectral representation of the Euclidean correlator at small momenta.
Finally, an analysis of the flavor structure of the electromagnetic current correlator
is performed, where a recent theoretical estimate of the Wick-disconnected diagram contributions 
is rederived independently and confirmed.
\end{abstract}

\maketitle

\section{Introduction}

The hadronic vacuum polarization, that is, the way hadrons modify the
propagation of virtual photons, is of great importance in precision
tests of the Standard Model of particle physics. It enters, for
instance, the running of the QED coupling constant. Together with the
Higgs mass, the latter can be used to predict the weak mixing angle,
which can also be measured directly, thus providing for a test.
Secondly, it currently represents the dominant uncertainty in the 
Standard Model prediction of the anomalous magnetic moment of the muon.
Given the upcoming experiment at FermiLab that is expected to improve
the accuracy of the direct measurement by a factor 4, it is important
to reduce the uncertainty on the prediction by a comparable factor.
While the phenomenological determination of the leading hadronic contribution
is still the most accurate approach, a purely theoretical prediction 
is both conceptually desirable and provides for a completely independent 
check. Since the vacuum polarization is inserted into an integral which is strongly
weighted to the low-energy domain, calculating the hadronic vacuum polarization
has become an important goal for several lattice QCD collaborations
performing non-perturbative simulations
\cite{Blum:2002ii,Gockeler:2003cw,Aubin:2006xv,Feng:2011zk,Boyle:2011hu,DellaMorte:2011aa}.

One of the features of the numerical lattice QCD framework is that the
theory is formulated in Euclidean space of finite extent.  Typically
the theory is set up on a four-dimensional torus.  The limitation of
Euclidean correlation functions in finite volume to discrete values of
the momenta has drawn considerable attention
recently~\cite{deDivitiis:2012vs,Aubin:2012me,Feng:2013xsa}. Many low-energy
quantities defined in infinite volume, such as the slope of the Adler
function at the origin or the proton radius defined from the slope of
its electric form factor at $Q^2=0$, do not have a unique, canonical
definition in finite volume. Instead, different finite-volume
representations can be defined, all of which converge to the desired
infinite-volume quantity. From the point of view of lattice QCD
simulations, a desirable feature of such a representation is that it
converges rapidly to the infinite-volume quantity. A different
representation can in general be obtained by deriving an equivalent formulation
of the infinite-volume quantity, and then carrying it over to the
finite-volume theory.

Even if a new representation provides a definition of the vacuum
polarization or a form factor for a continuous set of momenta, clearly
it only represents progress if the finite-size effect on the final
target quantity is reduced. Therefore the merit of a new representation
can only be evaluated once some theoretical understanding of the
finite-size effects is reached.

Here we explore a representation of the hadronic vacuum polarization
based on the time-momentum representation of the vector correlator.
The starting point is equation (\ref{eq:PihatGt}), which was
previously derived in~\cite{Bernecker:2011gh}. It suggests a way to
compute the hadronic vacuum polarization for any value of the
virtuality. In this paper we apply the idea in a lattice QCD
calculation with two light quark flavors.  From the appearance of a
power of the time-coordinate in the integral, it is manifest that a
derivative with respect to the Euclidean frequency has been
taken. With a finite and periodic time extent $T$, the function
$x_0^2$ is not uniquely defined. However, since it is multiplied by a
vector correlation function, which falls off exponentially, the
ambiguity is parametrically small. How precisely we deal with this
issue is presented in section \ref{sec:numerics}.

We address the finite-size effects on our representation of the Adler
function in section \ref{sec:fv}. We use the operator product
expansion to analyze the finite-size effects at large $Q^2$, and we
use the known connection between the finite-volume and the
infinite-volume spectral function at low energies to study the
finite-size effects at small virtualities. Even if our analysis does
not apply to intermediate distances, we expect the finite-size
effect coming from the long-distance part of the correlator to be the
dominant one. Our results suggest that the slope of the Adler function
at the origin is approached from below in the large volume limit.

An important feature of our method is that it applies irrespective of
the flavor structure of the current. The method of partially twisted
boundary conditions has so far been limited to isovector quantities
(see for instance \cite{DellaMorte:2011aa}). Here we present a lattice
calculation of the isovector contribution, which allows us to compare
our results with those obtained by the commonly used momentum-space method
on the same ensemble.  The isovector correlator does not require the
calculation of Wick-disconnected diagrams, whose standard estimators
are affected by a large statistical variance.  Recently, a useful
estimate of the size of the latter was derived in chiral perturbation
theory~\cite{Juttner:2009yb,DellaMorte:2010aq}.  Here we revisit this
relation, and show that it can be understood in terms of the higher
threshold at which the isosinglet channel opens compared to the
isovector channel.

During the final stages of this work, a preprint by Feng et
al.\ \cite{Feng:2013xsa} appeared with which the present paper has an
overlap. The authors of~\cite{Feng:2013xsa} emphasized the attractive
option of accessing the hadronic vacuum polarization at small momenta
with a very similar method.  They also explored the interesting
possibility of analytically continuing the vacuum polarization
function into the timelike region below threshold.

The structure of this paper is as follows.
Our definitions are collected in the next section.
The finite-volume effects are analyzed in section \ref{sec:fv},
with technical details to be found in the appendix.
The numerical calculation is described in section \ref{sec:numerics} and 
the results are given in section \ref{sec:results}.

\renewcommand{\vec}[1]{\boldsymbol{#1}}

\section{Definitions}

In this section we consider QCD in infinite Euclidean space.
The vector current is defined as $j_\mu(x)=\bar\psi(x)\gamma_\mu\psi(x)$, where the Dirac matrices
are all hermitian and satisfy $\{\gamma_\mu,\gamma_\nu\}=2\delta_{\mu\nu}$. 
The flavor structure of the current will be discussed in the next subsection. We use 
capital letters for Euclidean four-momenta and lower-case letters for Minkowskian four-momenta.
In Minkowski space we choose the `mostly-minus' metric convention.
In Euclidean space, the natural object is the polarization tensor
\be\la{eq:PolTens}
\Pi_{\mu\nu}(Q) \equiv \int d^4x \, e^{iQ\cdot x} \<j_\mu(x) j_\nu(0)\>,
\ee
and O(4) invariance and current conservation imply the tensor structure
\be\la{eq:PimunuQ}
\Pi_{\mu\nu}(Q) = \big(Q_\mu Q_\nu -\delta_{\mu\nu}Q^2\big) \Pi(Q^2).
\ee
With these conventions, the spectral function
\be\la{eq:rhoGG}
\rho(q^2) \equiv -\frac{1}{\pi} {\rm Im} \Pi(Q^2)\Big|_{{Q_0=-iq_0+\epsilon},\,{\vec Q=\vec q}} 
\ee
is non-negative for a flavor-diagonal correlator. 
For the electromagnetic current, it is related to the $R$ ratio via
\be \la{eq:rhoR}
\rho(s) =\frac{R(s)}{12\pi^2},
\qquad
R(s) \equiv  \frac{\sigma(e^+e^-\to {\rm hadrons})}
 {4\pi \alpha(s)^2 / (3s) } .
\ee
The denominator is the treelevel cross-section $\sigma(e^+e^-\to\mu^+\mu^-)$
in the limit $s\gg m_\mu^2$, and we have neglected QED corrections.

Relation (\ref{eq:rhoGG}) can be inverted. 
The Euclidean correlator is recovered through a dispersion relation,
\be \la{eq:DispRel}
\frac{\widehat\Pi(Q^2)}{4\pi^2}\equiv \Pi(Q^2)-\Pi(0) = Q^2 \int_0^\infty ds \frac{\rho(s)}{s(s+Q^2)}.
\ee
Finally we introduce the mixed-representation Euclidean correlator
\be\la{eq:Gdef}
G(x_0)\delta_{k\ell} = - \int d^3\vec x\; \< j_k(x) j_\ell(0) \>,
\ee
which has the spectral representation~\cite{Bernecker:2011gh}
\be\la{eq:Gspecrep}
G(x_0) = \int_0^\infty d\omega\; \omega^2\rho(\omega^2)\; e^{-\omega |x_0|},
\qquad x_0\neq 0.
\ee
The vacuum polarization can be expressed as an integral over $G(x_0)$~\cite{Bernecker:2011gh},
\ba  \la{eq:PihatGt}
&&\Pi(Q_0^2)-\Pi(0)  =
\\ && \qquad  \int_{0}^\infty \!\!\! dx_0\, G(x_0)\Big[x_0^2 - \frac{4}{Q_0^2}\sin^2(\half Q_0 x_0) \Big].
\nonumber
\ea
From here the Adler function is given by
\ba\la{eq:Adler}
&& D(Q_0^2) \equiv 12\pi^2 Q_0^2 \frac{d\,\Pi}{dQ_0^2} 
 = \frac{12\pi^2}{Q_0^2} \int_0^\infty dx_0 \, G(x_0) 
\\ && \quad \qquad \qquad \left(2-2\cos(Q_0 x_0) - Q_0x_0\sin(Q_0x_0)\right).
\nonumber
\qquad 
\ea
The slope of the Adler function at the origin is of particular interest,
\be\la{eq:Dp0}
D'(0)=\lim_{{Q^2\to0}}\frac{D(Q^2)}{Q^2} = \pi^2 \int_0^\infty dx_0 \;x_0^4\, G(x_0).
\ee
For instance, in the case of the electromagnetic current,
 the hadronic contribution $a_\ell^{\rm HLO}$ to the anomalous magnetic moment of a lepton is given,
in the limit of vanishing lepton mass, by~\cite{Bernecker:2011gh}
\be
\lim_{m_\ell\rightarrow 0}\frac{a_\ell^{\rm HLO}}{m_\ell^2} = 
\frac{1}{9}\Big(\frac{\alpha}{\pi}\Big)^2 D'(0).
\ee

\subsection{A note on flavor structure in the $N_{\rm f}=2$ theory}

For simplicity we consider isospin-symmetric two-flavor QCD.
The electromagnetic current is then given by 
$j^{\gamma}_\mu = j^\rho_\mu + {\txts\frac{1}{3}} j^\omega_\mu$ with
\be\la{eq:jmurhodef}
j_\mu^\rho \equiv \frac{1}{2} (\bar u \gamma_\mu u - \bar d \gamma_\mu d),\qquad 
j_\mu^\omega \equiv \frac{1}{2} (\bar u\gamma_\mu u + \bar d\gamma_\mu d).
\ee
For each of these currents $f=\rho,\omega,\gamma$, 
we define polarization tensors $\Pi_{\mu\nu}^{ff'}$ as in \eq(\ref{eq:PolTens}).
Obviously only two are linearly independent, and in particular
\be
\Pi^{\gamma\gamma}_{\mu\nu}(q)
= \Pi^{\rho\rho}_{\mu\nu}(q) + \frac{1}{9}\Pi^{\omega\omega}_{\mu\nu}(q)  .
\ee

A very interesting relation was recently
obtained~\cite{Juttner:2009yb} between the contributions of the
Wick-connected diagrams and the Wick-disconnected diagrams in
$\Pi^{\gamma\gamma}_{\mu\nu}(q)$, similar to \eq(\ref{eq:1ov10})
below.  The derivation was based on an NLO calculation in chiral
perturbation theory (ChPT), and extended to include the strange
quark~\cite{DellaMorte:2010aq}.  Here we rederive the result in a
different way without relying on ChPT.  In terms of Wick contractions,
the Euclidean correlators are given by
\ba\la{eq:rhowcwd}
\Pi^{\rho\rho}_{\mu\nu}(q) &=&  \frac{1}{2} \Pi^{\rm wc}_{\mu\nu}(q),
\\  \la{eq:omwcwd}
\Pi^{\omega\omega}_{\mu\nu}(q) &=&  \frac{1}{2} \Pi^{\rm wc}_{\mu\nu}(q) + \Pi^{\rm wd}_{\mu\nu}(q),
\\ \la{eq:gwcwd}
\Pi^{\gamma\gamma}_{\mu\nu}(q) &=&  \frac{5}{9} \Pi^{\rm wc}_{\mu\nu}(q) + \frac{1}{9}\Pi^{\rm wd}_{\mu\nu}(q),
\ea
where `wc' and `wd' stand for Wick-connected and Wick-disconnected diagrams respectively.

By linearity, spectral functions corresponding to $\Pi^{\rm  wc}_{\mu\nu}$ and $\Pi^{\rm wd}_{\mu\nu}$ can be defined as in
\eq(\ref{eq:PimunuQ}, \ref{eq:rhoGG}), although $\rho^{\rm  wd}(s)$ is then not necessarily positive definite.  In the
isovector channel, the threshold opens at $\sqrt{s}=2m_\pi$, therefore by \eq(\ref{eq:rhowcwd}),
$\rho^{\rm wc}(s)$ becomes non-zero at the same center-of-mass
energy. In the isosinglet channel it opens at $\sqrt{s}=3m_\pi$,
\be
\rho^{\omega\omega}(s) = 0, \qquad 0< \sqrt{s} < 3m_\pi.
\ee
In terms of the Wick contractions, this means, from \eq(\ref{eq:omwcwd})
\be\la{eq:wdwc}
\rho^{\rm wd}(s) = -\frac{1}{2} \rho^{\rm wc}(s),\qquad \sqrt{s} < 3m_\pi.
\ee
In particular, from \eq(\ref{eq:gwcwd}), the contribution of the Wick-disconnected contribution to the Wick-connected
contribution in the electromagnetic current spectral function is given by
\be\la{eq:1ov10}
\frac{\frac{1}{9}\rho^{\rm wd}(s)}{\frac{5}{9} \rho^{\rm wc}(s)} = -\frac{1}{10},
\qquad 2m_\pi < \sqrt{s} <3m_\pi.
\ee
This result is exact in two-flavor QCD with isospin symmetry.  The
derivation shows that it stems essentially from the higher energy
threshold at which it becomes possible to produce an isosinglet
state. Because experimental $e^+e^-$ data shows that the three-pion
channel opens rather slowly (the $\omega$ resonance is very narrow), 
relation (\ref{eq:1ov10}) can be expected to be a good
approximation at least up to 700MeV.  For instance the contribution to the $R$
ratio of the $\pi^+\pi^-\pi^0$ channel to the $R(s)$ ratio is of order
0.01 at $\sqrt{s}=700{\rm MeV}$~\cite{Dolinsky:1991vq}, while the $R(s)$ ratio itself lies
between 4.0 and 5.0 at the same center-of-mass energy. The smallness of the ratio (\ref{eq:1ov10})
stems mainly from the small charge factor multiplying the Wick-disconnected contribution in (\ref{eq:gwcwd}).

The relation (\ref{eq:wdwc}) between the Wick-disconnected and the
Wick-connected contribution can be translated back into the Euclidean
correlator via the dispersion relation (\ref{eq:DispRel}). A stronger statement can be
made in the time-momentum representation (\ref{eq:Gdef}), since the low-energy part of
the spectral function dominates exponentially at large Euclidean time
separations. 
For $x_0\to\infty$ we have
\be
G^{\rm wd}(x_0) = -\frac{1}{2} G^{\rm wc}(x_0)\; \left( 1 + {\rm O}(e^{-m_\pi x_0})\right).
\ee
Unlike at short distances, where $G^{\rm wd}/G^{\rm wc}$ is of order
$\alpha_s^3$~\cite{Baikov:2012zm}, the Wick-disconnected diagram is
thus of the same order as the Wick-connected diagram at long
distances. 

The argument just presented can be made for other symmetry channels and can be extended
to include the other quark flavors.

\section{Finite-size effects on the Adler function\la{sec:fv}}

In this section we investigate the finite-size effects on the Adler function
specifically for the representation (\ref{eq:Adler}), although the methods
used are more generally applicable. 

\subsection{Large momenta}

We denote by $\Pi_{\mu\nu}(Q,L,T)$ the polarization tensor on an
$T\times L^3$ torus to distinguish it from its infinite-volume
counterpart $\Pi_{\mu\nu}(Q)$, and by $\Delta\Pi_{\mu\nu}(Q,L,T)\equiv
\Pi_{\mu\nu}(Q,L,T)-\Pi_{\mu\nu}(Q)$ the finite-size effect (we will
use the same notational convention for other quantities). Although
the polarization tensor itself contains a logarithmic ultraviolet
divergence, its finite-size effect is ultraviolet finite.  When
discussing finite-size effects, the specific finite-volume
representation used must be specified~\cite{Bernecker:2011gh}. Consider then the
Fourier transform of $G(x_0)$. At large frequency, its finite-size effect
is given by the operator product expansion,
\ba\la{eq:Pikk}
&&\frac{1}{3}\Delta\Pi_{kk}((Q_0,\vec 0),L,T) =\frac{1}{Q_0^2} \sum_{i=1}^4 C_i(Q_0^2,\mu^2) 
\\ &&~~~~ \left(\<O_4^{(i)}(\mu)\>_{(L,T)} - \<O_4^{(i)}(\mu)\>_{\infty}\right) + {\rm O}(1/Q_0^4). \nonumber
\ea
Dimension-four operators that contribute are the Lorentz scalar,
renormalization group invariant operators $\frac{\beta(g)}{2g}G^2$ and
$m\bar\psi\psi$, but also the $(00)$ component of the two
flavor-singlet, twist-two, dimension-four operators familiar from deep
inelastic scattering.  The $Q^2$ dependence of the Wilson coefficients
$C_i(Q^2,\mu^2)$ is logarithmic. The coefficients of the Lorentz
scalar operators are known to next-to-leading
order~\cite{Chetyrkin:1985kn} while the coefficients of the traceless
tensor operators are known at leading order~\cite{Mallik:1997pq}.  The
latter can be taken from the calculation \cite{Mallik:1997pq} performed for 
thermal field theory in infinite volume, since on the $T\times L^3$ torus, 
the expectation value of 
a traceless rank-two tensor operator only has one independent non-vanishing component
in spite of the lack of rotational invariance in a time slice.

We thus see that the relative volume effect on the
polarization tensor is suppressed by a factor $Q^4$. Furthermore, the
finite-size effect on the expectation value of a local operator
$O_4^{(i)}(\mu)$ appearing in \eq(\ref{eq:Pikk}) is of order $e^{-m_\pi L}$ for
sufficiently large $L$. This fact is familiar from finite-temperature
QCD. The leading finite-size effect is due to a one-particle state, so
that the prefactor of $e^{-m_\pi L}$ can be related to a pion matrix
element~\cite{Meyer:2009kn}.

The lesson is that for momenta sufficiently large that the
Fourier-transformed product of currents can be represented by a
\emph{local} operator, the asymptotic finite-size effects on the
polarization tensor are of order $e^{-m_\pi L}$.  This statement about
the finite-size effect on $\Pi_{\mu\nu}$ then carries over to the
Adler function at large $Q^2$.

\subsection{Long-distance contribution to $D'(0)$}

\begin{figure}[t]
\includegraphics[width=.5\textwidth]{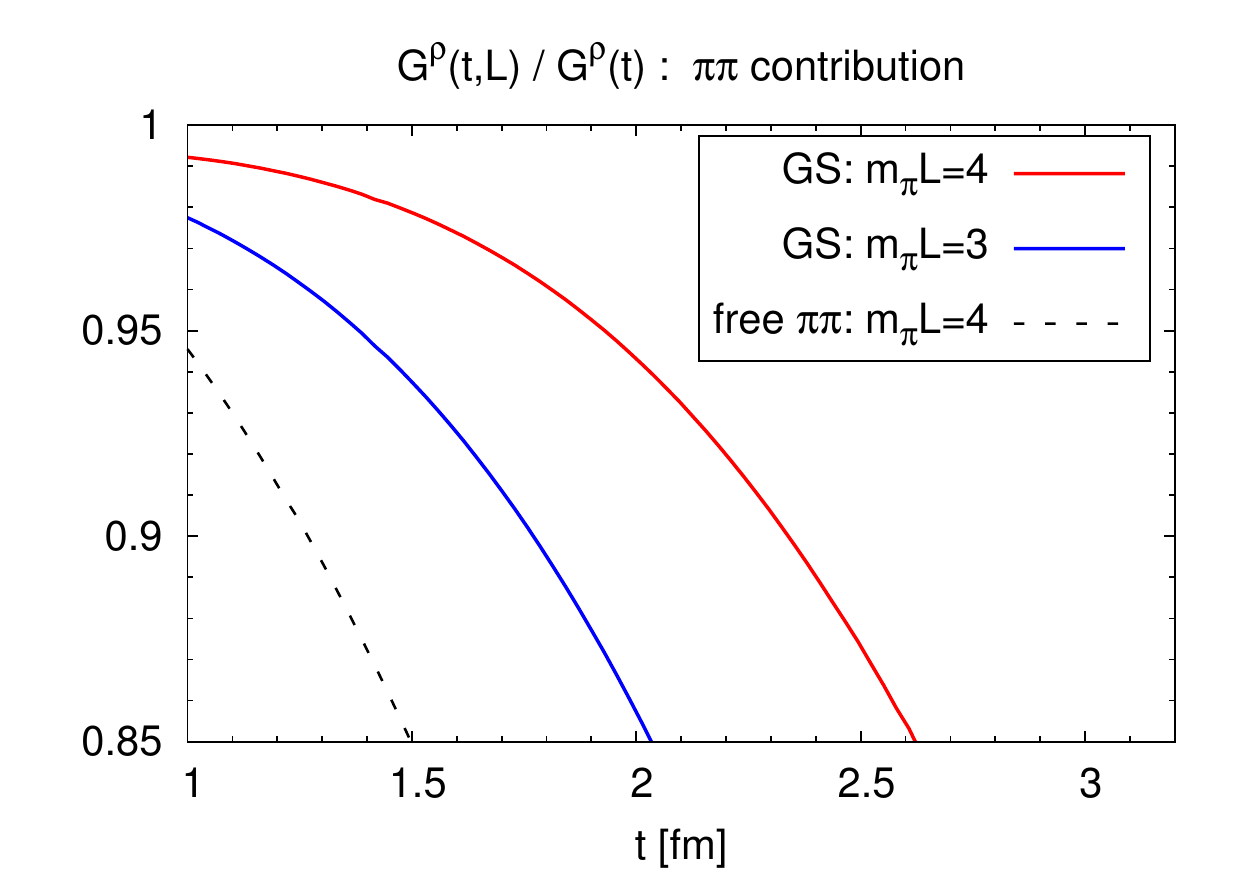}
\caption{Relative finite-size effect on the Euclidean correlator (\ref{eq:Gdef})
for a pion form factor given by the Gounaris-Sakurai (GS) parametrization
and for free pions. For details see appendix~A.}
\label{fig:Gtrat}
\end{figure}

One of the most important observables is the slope $D'(0)$ of the
Adler function at $Q^2=0$ (which up to a numerical factor coincides
with the slope of the vacuum polarization).  It determines the leading
hadronic contribution to the anomalous magnetic moment of the
electron~\cite{Bernecker:2011gh}, and a large fraction of the muon's
anomalous magnetic moment~\cite{Brandt:2012zza}. It is theoretically
attractive, because it involves no energy scale external to QCD.  In
the representation given in \eq(\ref{eq:Dp0}), the dominant
contribution comes from Euclidean time separations $x_0>1{\rm fm}$. We
therefore find it useful to define
\be\la{eq:Dpeff}
D'_{t} \equiv \pi^2 \int_{t}^\infty dx_0\, x_0^4 \, G(x_0).
\ee
so that $D'_0 = D'(0)$.  In the infinite-volume theory, the
contribution of the $|\pi^+\pi^-\>_{\rm out}$ states with an energy up
to and including the $\rho$ mass completely dominates this
contribution. Given the argument made above on the finite-size effects
on the short-distance contribution to the Adler function, and the form
of the finite-size effects in free field theory (see section
\ref{sec:freep}), we expect the finite-size effects on $D'(0)$ to be
dominated by the finite-size effects on $D'_t$ for $t={1{\rm fm}}$. We
are therefore led to discuss the latter. We will only consider the
case where the pion mass is set to its physical value, but with a
suitable model for the timelike pion form factor $F_\pi(\omega)$, 
our analysis can be extended to other pion masses.

If a temporal extent $T=2L$ is chosen, as is common practice, the
dominant finite-size effect comes from the finite spatial box extent.
In the following theoretical analysis we therefore set $T$ to
infinity.  How to proceed in practice where $T$ is finite is discussed
in section (\ref{sec:results}).

\begin{table}[t]
\begin{tabular}{l|r|r|r}
 $k/m_\pi$    &  $~|A|^2/m_\pi^3~$   & $~~~\mathbb{L}(k)~~~$ &  $~|F_\pi(\omega)|^2$\\
\hline
 1.548   & 0.0737   & 6.606   & 3.507\\
 2.133   & 0.4702   & 11.53   & 12.85\\
 2.559   & 1.1333   & 28.95   & 42.52\\
 2.831   & 0.7509   & 23.09   & 16.21\\
 3.171   & 0.1124   & 62.41   & 4.558\\
 3.581   & 0.1452   & 25.13   & 1.615\\
 3.912   & 0.1335   & 19.36   & 0.867\\
 4.459   & 0.0192   & 91.10   & 0.391\\
\end{tabular}
\caption{The first eight energy levels and matrix elements
on a torus of linear size $L=4/m_\pi$ for the Gounaris-Sakurai parametrization of
$F_\pi(\omega)$ (see appendix A; $\omega \equiv 2\sqrt{m_\pi^2 + k^2}$). The `Lellouch-L\"uscher' 
factor $\mathbb{L}(k)$ and the pion form factor
at the corresponding energies are given in columns 3 and 4.}
\label{tab:states}
\end{table}

One source of finite-size effects are the polarization effects on
single-particle states.  They have been analyzed in detail in the
past~\cite{Luscher:1985dn}. The upshot is that the properties of these
states are only affected by corrections that are exponential in the
linear torus size.  In the following, we will neglect these
finite-size corrections. In lattice QCD calculations, this assumption
will have to be checked explicitly.

In~\cite{Bernecker:2011gh}, an analysis of the finite-size effects on
$G(x_0)$ was carried out using the relation between the finite-volume
spectral function and the infinite-volume spectral
function~\cite{Luscher:1991cf,Meyer:2011um}.  This relation is only
firmly established up to the inelastic threshold of $\omega=4m_\pi$.
Here we will be less rigorous and assume that even somewhat above this
threshold, the relation remains a good approximation. The main
justification for this assumption is that the $\rho$ decays almost
exclusively into two pions.  We also neglect possible contributions
from $\pi\pi$ scattering in the $\ell=3$ and higher partial waves.

Before using the full machinery of L\"uscher's finite-volume formalism,
it is worth understanding the qualitative behavior of $\Delta D'_t$
in two simple, opposite limits. In one limit, we have non-interacting pions,
$F_\pi(\omega)=1$ and $G(x_0)$ can be computed exactly both in finite and 
in infinite volume. The ratio of the finite-volume to the infinite-volume
correlation function is diplayed in \fig(\ref{fig:Gtrat}). Clearly 
the finite-size effects are large for a typical value of $m_\pi L=4$.

In the other limit, the pion interactions are such that the vector current
only couples to a stable $\rho$ meson, $\rho(s) = F_\rho^2 \delta(s-m_\rho^2)$.
This limit correspond to the expected behavior at very large number of colors $N_c$.
In this case, $ G(x_0,L) = \frac{1}{2} F_\rho^2 m_\rho e^{-m_\rho x_0}$.
The only finite-volume effect in this case stems from the finite-size 
effects on $m_\rho$ and $F_\rho$, which are exponentially small in the volume.
Thus in this limit the finite-size effects are expected to be much more benign.

Experimental $e^+e^-$ data shows that QCD lies somewhere in between
these two extremes, but is somewhat closer to the narrow resonance
limit.  Using the Gounaris-Sakurai parametrization of the timelike
pion form factor~\cite{Gounaris:1968mw}, we have calculated the eight
lowest energy-eigenstates for $m_\pi L=3$ and $4$ and their coupling
to the isospin current using the results
of~\cite{Luscher:1991cf,Meyer:2011um}, see Table \ref{tab:states}. For
details of the calculation we refer the reader to appendix A.
Together these states saturate the correlator beyond 1fm to a high
degree of accuracy.  The ratio $G(x_0,L)/G(x_0)$ is displayed in
\fig\ref{fig:Gtrat}.  Although the relative finite-size effect grows
rapidly at long distances, for $m_\pi L=4$ it is of acceptable size at
the distances which dominate $D'(0)$ (compare with
Fig.\ \ref{fig:slope}).  Finally, table \ref{tab:Dp0} gives the
finite-size effect on $D_t'$ for $t=1{\rm fm}$.  With $m_\pi L=4$, the
effect amounts to $7\%$, which is a larger effect than is observed for
many mesonic observables. We also note that at the heavier quark
masses currently studied in the simulations, the $\rho$ is presumably
narrower and the finite-size effect for the same value of $m_\pi L$ is
therefore smaller.

\begin{table}[t]
\begin{tabular}{c|c|c}
$m_\pi L$~ &   ~~~GS~~~  & ~free pions \\
\hline
3 & 0.853  & 0.429 \\
4 & 0.927  & 0.671 \\
5 & 0.961  & 0.828 \\
\end{tabular}
\caption{The relative finite-size effect $D'_t(L)/D_t'$ for  $t=1{\rm fm}$
(see \eq(\ref{eq:Dpeff}) and (\ref{eq:Adler})).
}\label{tab:Dp0}
\end{table}

\begin{figure}[t]
\includegraphics[width=.5\textwidth]{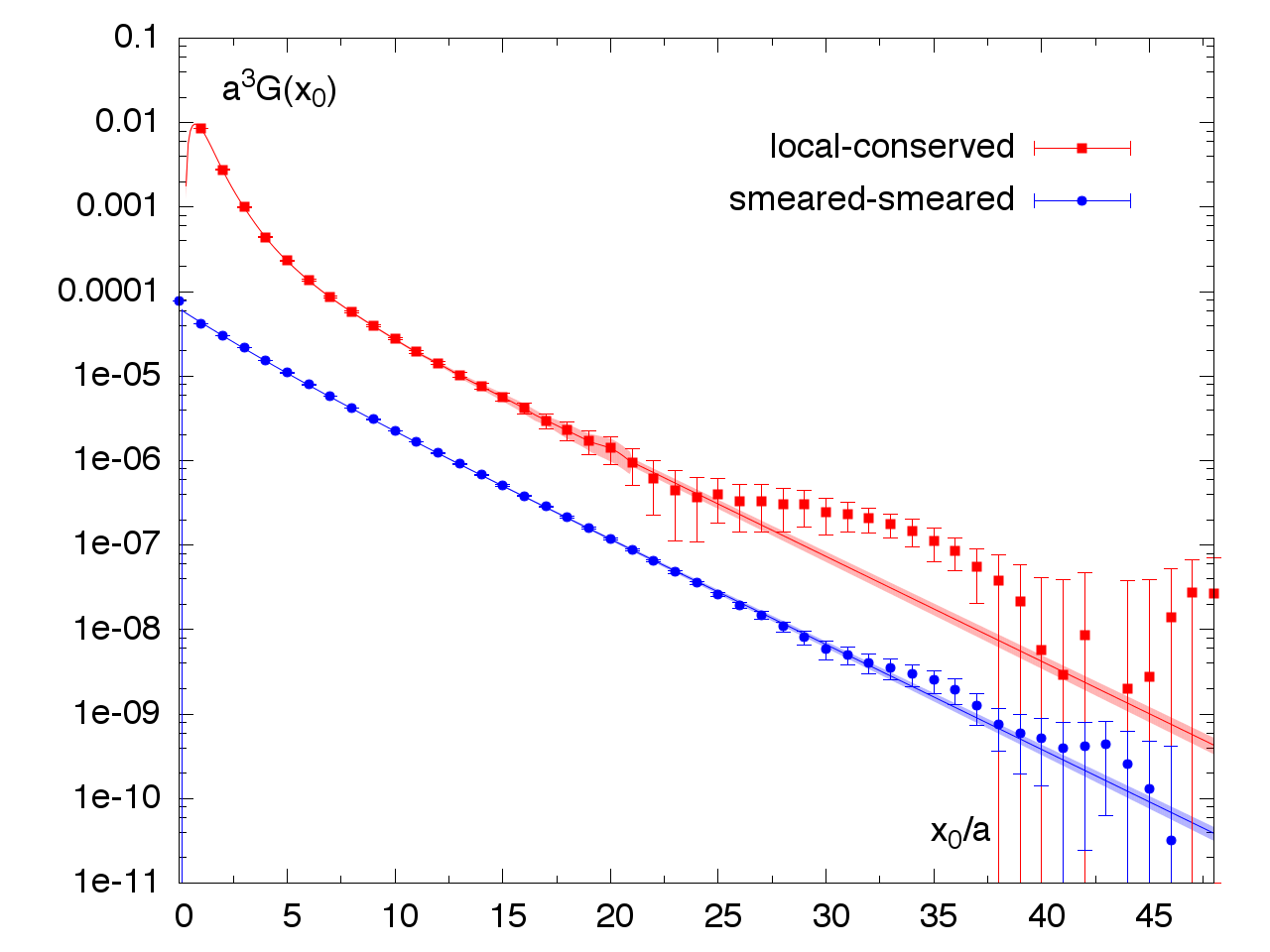}
\caption{{Local-conserved and smeared-smeared isovector vector correlation functions. 
The red shaded area shows the correlator entering Eq.\ (\ref{eq:PihatGt})
for the computation of $\hat\Pi(Q^2)$, the blue shaded area correlator is used 
to fit the lowest lying mass for extrapolation to all time beyond $x_0\simeq T/4$.}}
\label{fig:corr}
\end{figure}

\section{Numerical Setup\la{sec:numerics}}

In this and the following sections, we describe a numerical
implementation of the representation of the vacuum polarization and
Adler function given in \eq(\ref{eq:PihatGt},~\ref{eq:Adler}). In
particular, we show how the integral over the time coordinate can be
treated without introducing unnecessarily large finite time-extent
effects. We restrict ourselves to one value of the light quark mass
and one lattice spacing for which we can directly compare the results
obtained with the new method to those obtained with the momentum-space
method~\cite{DellaMorte:2012cf}.

All our numerical results were computed on dynamical gauge
configurations with two mass-degenerate quark flavors.  The gauge
action is the standard Wilson plaquette action \cite{Wilson:1974sk},
while the fermions were implemented via the O($a$) improved Wilson
discretization with non-perturbatively determined clover coefficient
$c_{\rm sw}$ \cite{Jansen:1998mx}.  The configurations were generated
using the DD-HMC algorithm~\cite{Luscher:2005rx,Luscher:2007es} as
implemented in L\"uscher's DD-HMC package~\cite{CLScode} and were made
available to us through the CLS effort~\cite{CLS}. We calculated
correlation functions using the same discretization and masses as in
the sea sector on a lattice of size $96\times 48^3$ (labeled F6
in~\cite{Capitani:2012gj}) with a lattice spacing of $a =
0.0631(21)$fm~\cite{Capitani:2011fg} and a pion mass of
$m_\pi=324$MeV, so that $m_\pi L = 5.0$.

Regarding the flavor structure, we restrict ourselves to the isovector current 
$j_\mu^\rho$, normalized as in \eq(\ref{eq:jmurhodef}). 
On the lattice we implement the correlation function (\ref{eq:Gdef})
as a mixed correlator between the local and the conserved current, 
\be\la{eq:Gdeflc}
G^{\rm bare}(x_0,g_0)\delta_{kl} = - a^3 \sum_{\vec x} \< J^c_k(x) J^l_\ell(0) \>,
\ee
where
\ba
J_\mu^l(x) &=& \bar q(x) \gamma_\mu q(x),
\\
\la{eq:Jdef}
J_\mu^c ( x ) &=& \frac{1}{2} \Big(\bar q ( x + a\hat\mu ) ( 1 + \gamma_\mu ) U_\mu^\dagger ( x ) q ( x ) \nonumber 
\\
&& - \bar q ( x )( 1 - \gamma_\mu ) U_\mu ( x )q ( x + a\hat\mu ) \Big).
\ea
We have renormalized the vector correlator using
\ba
G(x_0)= Z_V(g_0)\;  G^{\rm bare}(x_0,g_0) 
\ea
with the non-perturbative value of
$Z_V=0.750(5)$~\cite{DellaMorte:2005rd}.  We have not included O($a$)
contributions from the improvement term proportional to the derivative
of the antisymmetric tensor
operator~\cite{Luscher:1996sc,Sint:1997jx}.  A quark-mass dependent
improvement term of the form $(1+b_V(g_0)am_q)$~\cite{Sint:1997jx} was
also neglected. These contributions should eventually be included to ensure a
smooth scaling behavior as the continuum limit is taken.  Here, our
primary goal is to test the method on a single ensemble.

\section{Numerical results\la{sec:results}}

We begin by analyzing results for the correlator $G(x_0)$, and then show
how the latter can be used to compute the subtracted vacuum polarization and
the Adler function.

\subsection{Correlator data}

\begin{figure}[t]
\includegraphics[width=.5\textwidth]{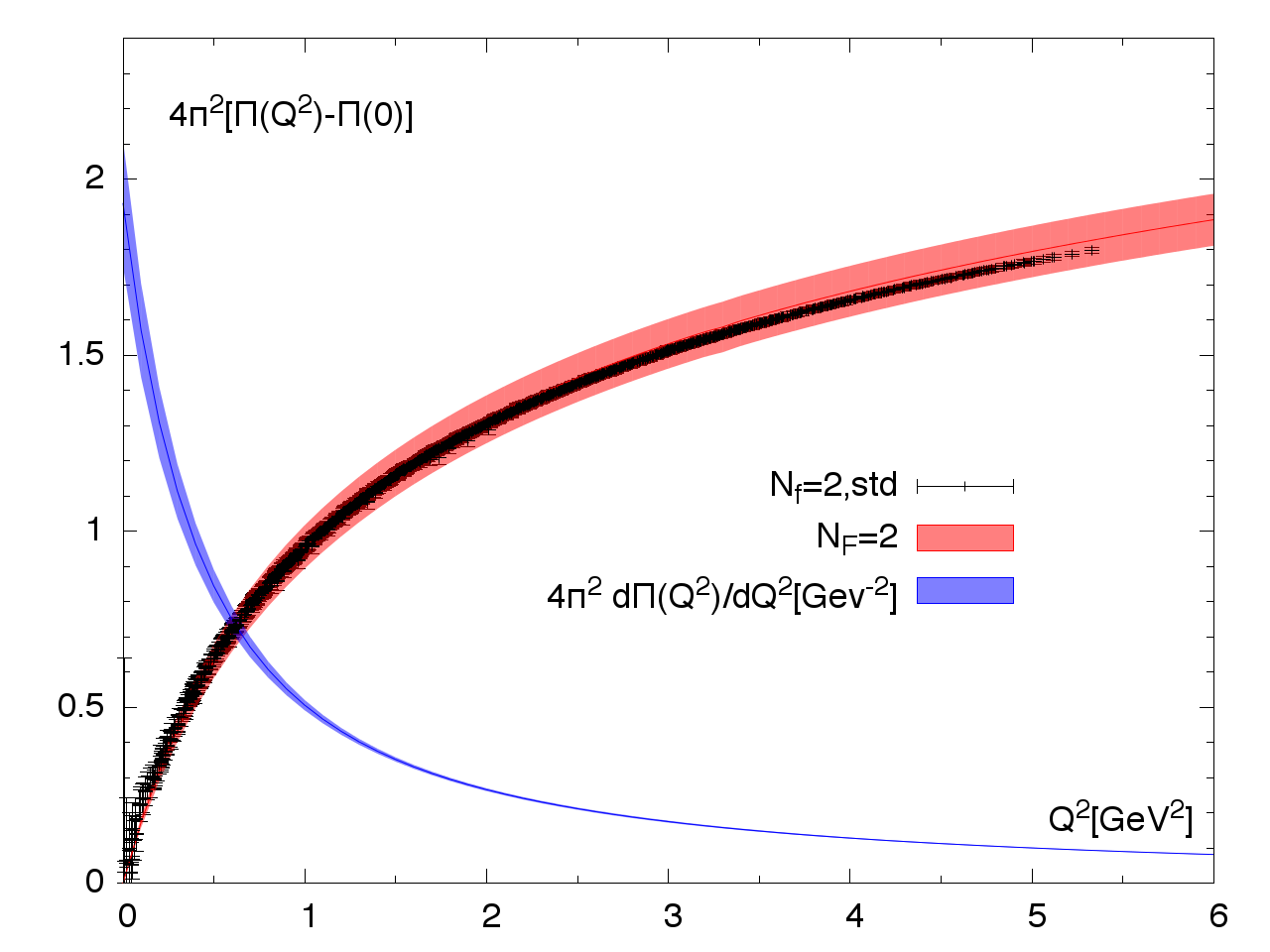}
\caption{{The subtracted vacuum polarization $\widehat\Pi(Q^2)$ and $d\widehat\Pi(Q^2)/dQ^2$ 
computed from the red shaded correlator in Fig.\ \ref{fig:corr}. 
The data shown in black were obtained using the momentum-space method 
on the same ensemble with comparable statistics \cite{DellaMorte:2012cf}.}}
\label{fig:pihat}
\end{figure}

In Fig.\ref{fig:corr} we show the local-conserved vector correlation
function. One virtue of this discretization is that in infinite
volume it leads to the property
\be\la{eq:sr}
\int_{-\infty}^{\infty} dx_0 \; G(x_0) =0.
\ee
The correlator  must  drop to negative
values for very small time separations in order to fulfill the above
identity. Indeed we observe this negative contact term for very small time
separations $|x_0|\leq a$, and \eq(\ref{eq:sr}) is satisfied by our data. 

The goal is to compute $\widehat\Pi(Q^2)$ and its derivative from the
lattice correlation function using the continuum relation
\eq(\ref{eq:PihatGt}). To achieve this one has to carry out the
integral over all time separations.  On the lattice this is 
not straightforward, since only a finite number of
points are available. In addition, the signal deteriorates rapidly
at large time separations.  It is known however that the correlation
function decays exponentially for large times. Therefore it is natural
to extrapolate the local-conserved correlator with an exponential that
decays with the lowest lying `mass'\footnote{The energy level extracted 
in this way does not necessarily correspond to a stable vector particle.}. 
This mass can be fixed by fitting the lattice data to an Ansatz of the form
\be
G_{\rm Ansatz}(x_0)=\sum_{n=1}^2 |A_n|^2 e^{-m_n x_0},
\ee
for $x_0$ sufficiently below $T/2$ that the `backward' propagating states 
make a negligible contribution. To ensure a reliable determination of this mass,
we have extracted it from a separate correlation function, computed on the 
same configurations using smeared operators at the source and sink~\cite{Capitani:2011fg}.
This correlator has greater overlap with the
ground state and yields very precise data. The mass parameter
determined in this way is then carried over to the local-conserved
correlator and the corresponding exponential is smoothly connected to
the lattice data by fitting $|A_1|^2$ to the data around $x_0=T/4$.

\begin{figure}[t]
\includegraphics[width=.5\textwidth]{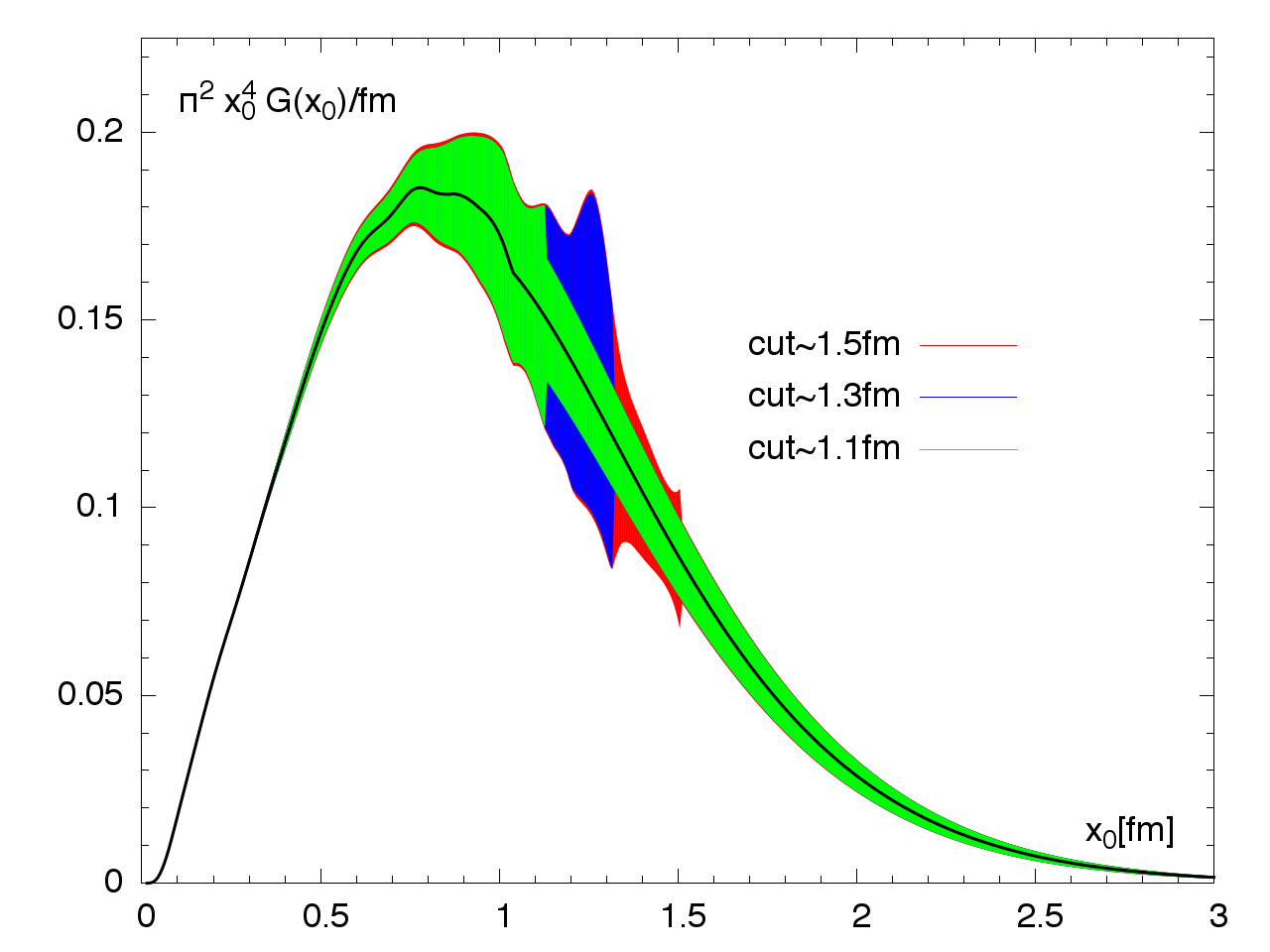}
\caption{{The integrand needed to compute the slope $D'(0)$ of the Adler function at $Q^2=0$. 
The bands of different color are the results obtained by replacing the data by 
a pure exponential falloff around the value $x_0=$cut[fm].
The area under this curve (divided by $3\cdot(0.197)^{2}$) is equal to
the intercept of the blue curve in \fig\ref{fig:pihat}.}}
\label{fig:slope}
\end{figure}

The resulting correlation function is shown as the red
shaded band in Fig.\ \ref{fig:corr}, where the error estimates were
obtained via a jackknife procedure. In the transition region from the
data dominated to the extrapolation dominated result the errors
increase for a small number of time steps, on the whole however
reasonably small errors are achieved in this way.

\subsection{Computing $\widehat\Pi(Q^2)$ and its slope}

In order to obtain $\widehat\Pi(Q^2)$ given the local-conserved
correlation function of Fig.\ \ref{fig:corr} one has to compute its
convolution with the kernel 
\be
K(x_0,Q_0)=x_0^2 - \frac{4\sin^2(Q_0  x_0/2)}{Q_0^2}.
\ee  
Using \eq(\ref{eq:PihatGt}) derivatives are
directly accessible, see for instance \eq(\ref{eq:Dp0}).

In Fig.\ \ref{fig:pihat} we show the result for $\widehat\Pi(Q^2)$ and
the slope $d\widehat\Pi(Q^2)/dQ^2$, as computed from the red shaded
correlator in Fig.\ \ref{fig:corr}. Here all errors were computed
using a jackknife method on a total of 392 measurements.  Turning
first to $\widehat\Pi(Q^2)$, for comparison we show the result
obtained on the same lattice using the standard method
\cite{DellaMorte:2012cf} with the same local-conserved discretization
and comparable statistics. The latter method consists in employing
Eqs.\ (\ref{eq:PolTens}, \ref{eq:PimunuQ}) to obtain first $\Pi(Q^2)$
and then determining $\Pi(0)$ via extrapolation. In this approach,
the number of data points at small $Q^2$ was significantly increased
using twisted-boundary conditions
\cite{deDivitiis:2004kq,Sachrajda:2004mi,Bedaque:2004ax}.  Still, the
extrapolation $\Pi(Q^2\rightarrow 0)$ is difficult to constrain as the
signal deteriorates in this limit.  The results obtained via
\eq(\ref{eq:PihatGt}) do not suffer directly from these issues, as the
physically relevant quantity, $\widehat\Pi(Q^2)$, is computed
directly.

Clearly the results obtained using our new method are very well
compatible with the standard method.  It should be noted that the
larger errors for large $Q^2$ only play a small role when computing
$a_\mu^{\rm HLO}$, as the large $Q^2$ region is highly suppressed in
the relevant integral. Using \eq(\ref{eq:Adler}) we also
display the slope $d\widehat\Pi(Q^2)/dQ^2$ as a blue shaded band in
Fig.\ref{fig:pihat}.  Throughout the result exhibits small statistical
errors and the intercept at $Q^2=0$ can be determined relatively
precisely.  We find $D'(0)=3\widehat\Pi'(0) = 5.8(5){\rm GeV}^{-2}$.

The factor $x_0^4$ in the integral representation \eq(\ref{eq:Dp0}) of the
derivative of the Adler function at the origin suppresses the small
time region of the correlator. The impact of each $x_0$-region can be
visualized by displaying the integrand, see Fig.\ \ref{fig:slope}.
Here we also show the results using three different values of the
transition point between the data and the extrapolation. This gives us
a handle to study the effect of the onset of the fitted pure
exponential described above.  The effect is seen to lie within the
error band, while the impact on the resulting $\widehat\Pi(Q^2)$ and
$d\widehat\Pi(Q^2)/dQ^2$ was checked and found to be negligible.
Examining the central value curve we observe that the dominant
contribution to the integrand is in fact given by the region
0.5fm$\leq x_0 \leq$1.5fm. Consequently, to precisely pin down $D'(0)$
and the closely related $a_\mu^{\rm HLO}$, very accurate lattice data
in this region is desirable.

\begin{figure}[t]
\includegraphics[width=.5\textwidth]{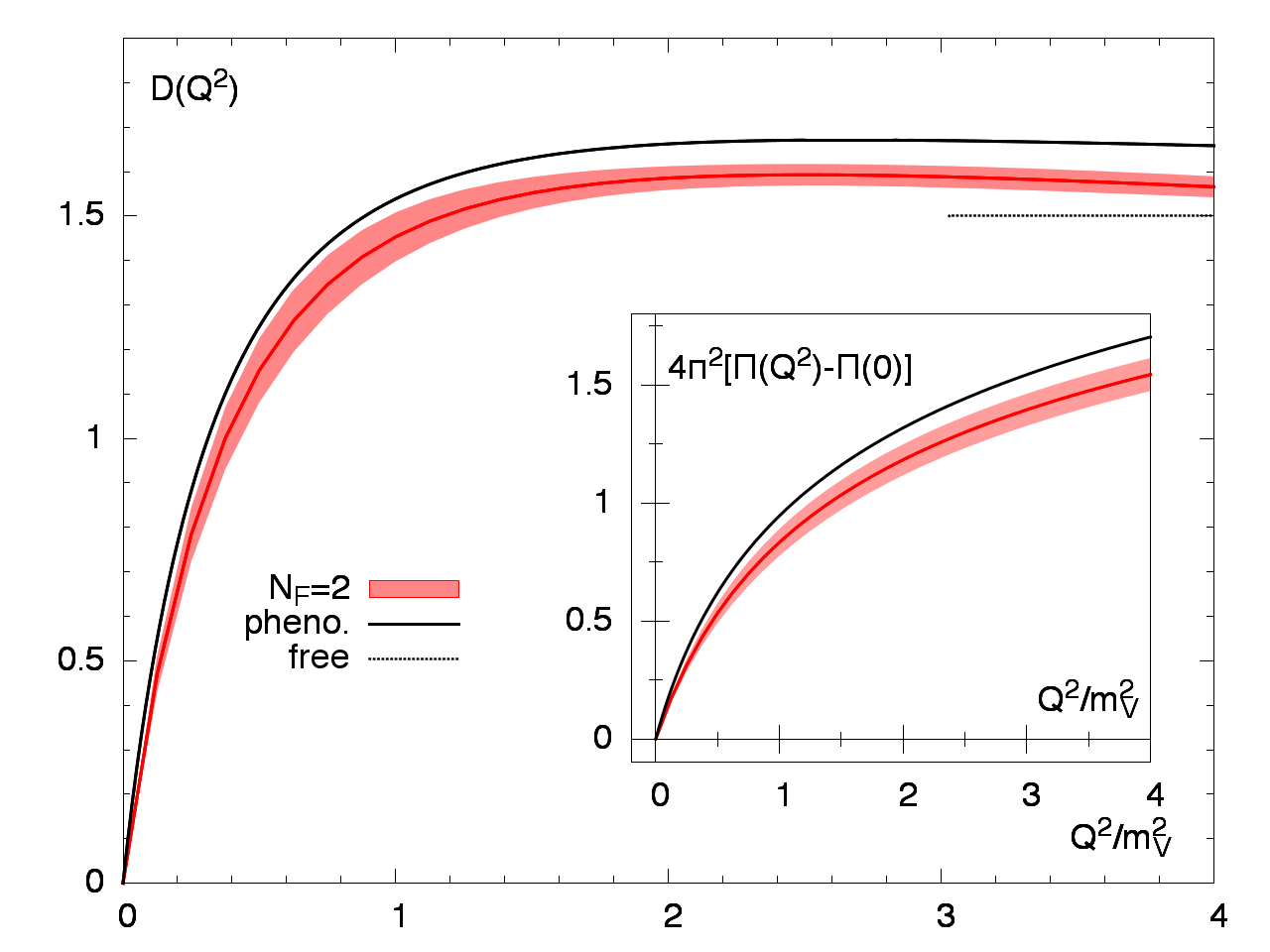}
\caption{{The functions $\widehat\Pi(Q^2)$ and $D(Q^2)$ from our analysis and 
a phenomenological model \cite{Bernecker:2011gh}. 
The horizontal axis has been rescaled by the ground-state mass $m_1=894(2)$MeV
on the lattice and the physical $\rho$ meson mass ($770$MeV) respectively.
For reference  the free result $D(Q^2)=\frac{3}{2}$ is shown as a dotted line.  }}
\label{fig:pihat2}
\end{figure}

In \fig\ref{fig:pihat2}, where we show the Adler function and the vacuum
polarization as a function of the virtuality, we follow the
approach~\cite{Feng:2011zk,Bernecker:2011gh} of rescaling the
horizontal axis using the vector meson mass. In this way one hopes to
achieve an approximate scaling at small $Q^2$, in the sense that the
curves corresponding to different quark masses approximately lie
on top of each other.  We compare our result to a phenomenological
model (Eq. (93) of \cite{Bernecker:2011gh}) for the isovector channel,
which predicts in particular $D'(0)=9.81(30){\rm GeV}^{-2}$.  The
comparison is shown in Fig.\ \ref{fig:pihat2} for the low $Q^2$ region
of $\widehat\Pi(Q^2)$ and the Adler function $D(Q^2)$.  Even after the
rescaling, the lattice data lie visibly below the phenomenological
curve. A plausible origin for the remaining difference is the
spectral density below the $\rho$ mass, since the integrand to obtain
$D'(0)$ is $R_1(s)/s^2$, where $R_1$ is the $R$-ratio restricted to
isovector final hadronic states.

\section{Conclusion}

We have tested a new representation of the vacuum polarization and the
Adler function $D(Q^2)$ which can be used in lattice QCD (see
also~\cite{Feng:2013xsa}). For the isovector contribution, we have
verified that it agrees well with the widely used method in
four-momentum space. In the latter case, we have
data~\cite{DellaMorte:2012cf} generated with twisted boundary
conditions, giving access to a discrete but dense set of
virtualities. By employing a representation that allows for continuous values
of the momenta, it is no more difficult to extract the Adler function
than the subtracted vacuum polarization.  The former has the advantage
of being local in $Q^2$, which facilitates the comparison with
perturbation theory at large $Q^2$.

A theoretical analysis of the finite-size effects associated with
$D'(0)$ suggests that the latter can be brought down to about five
percent at the physical pion mass for spatial volumes $m_\pi L$
between 4 and 5. The infinite-volume quantity is approached from
below.  At fixed $m_\pi L$ the finite-size effect depends strongly on
the width of the $\rho$ meson, implying that it rapidly becomes a more
critical issue when the pion mass is lowered towards its physical
value. If the masses and couplings of the low-lying vector states can
be determined on the lattice, the bulk of the finite-size effect can
be corrected for~\cite{Bernecker:2011gh}.

In the near future we plan to combine the extensive set of data that
we have generated with the standard method with the method presented
here to extract the Adler function at the origin and $a_{\mu}^{\rm
  HLO}$.  Subsequently, the calculation of the Wick-disconnected
diagrams can be taken up with our new representation.


%

\acknowledgments{
We are grateful to Michele Della Morte and Andreas J\"uttner, whose original
code formed the basis for our analysis programs, and 
to our colleagues within CLS for sharing the lattice ensemble used.  
We thank Georg von Hippel for discussions and for providing the smeared 
vector correlator~\cite{Capitani:2011fg}.
The correlation functions were computed 
on the dedicated QCD platform ``Wilson'' at the Institute for Nuclear Physics,
University of Mainz. 
This work was supported by the \emph{Center for Computational Sciences}
as part of the Rhineland-Palatinate Research Initiative.
}

\appendix

\section{Finite-size effects on the Euclidean correlator}

In this appendix we present the details of the calculation that underlies 
Tables (\ref{tab:states}, \ref{tab:Dp0}) and Figure (\ref{fig:Gtrat}).
It is based on the two-pion contribution to the spectral function.

The $\pi\pi$ contribution to the spectral function is given by 
(see for instance~\cite{Jegerlehner:2009ry})
\be
\rho(\omega^2) = \frac{1}{48\pi^2} 
\Big(1-\frac{4m_\pi^2}{\omega^2} \Big)^{\frac{3}{2}}  |F_\pi(\omega)|^2.
\la{eq:RFpi}
\ee
Charge conservation implies that $F_\pi(0)=1$.  Above the threshold
$\omega=2m_\pi$, the phase of the pion form factor is equal to the
$p$-wave pion phase shift, $F_\pi(\omega) = |F_\pi(\omega)|
e^{i\delta_{11}(k)}$ (Watson theorem).

\subsection{Interacting pions}
In infinite volume, the Euclidean correlation function is
obtained using \eq(\ref{eq:Gspecrep}, \ref{eq:RFpi}).
For the finite-volume correlator, we proceed as follows.

The discrete energy levels in the box and the infinite-volume phase
shifts are related by~\cite{Luscher:1990ux,Luscher:1991cf}
\ba \la{eq:Luscher_cond}
&&\delta_{11}(k) + \phi\left({\txts\frac{kL}{2\pi}}\right) = n\pi, 
\qquad n=1,2,\dots,
\\ &&\omega \equiv 2\sqrt{m_\pi^2 + k^2}.
\ea
The function $\phi(z)$, tabulated in \cite{Luscher:1991cf}, 
is defined by $\tan\phi(z) =
-\frac{\pi^{{3}/{2}}z}{{\cal Z}_{00}(1;z^2)}$, where ${\cal
  Z}_{00}(1;z^2)$ is the analytic continuation in $s$ of ${\cal
  Z}_{00}(s;z^2) = \frac{1}{\sqrt{4\pi}} \sum_{\boldsymbol
  n\in\mathbb{Z}^3} \frac{1}{(\boldsymbol{n}^2-z^2)^s}$.
The corresponding finite volume matrix elements for unit-normalized
finite-volume states are given by~\cite{Meyer:2011um}
\ba\la{eq:result1}
|F_\pi(\omega)|^2 &=& \mathbb{L}(k)
\frac{3\pi \omega^2}{2 k^5}    |A|^2 ,
\\
\mathbb{L}(k)&\equiv& \left[z \phi'(z)\right]_{z=\frac{kL}{2\pi}}
 + k \frac{\partial\delta_1(k)}{\partial k}.
\ea
The correlation function is then obtained as
\be
G(x_0,L) = \sum_n |A_n|^2 \; e^{-\omega_n x_0}.
\ee
We thus only need a realistic model for the timelike pion form factor $F_\pi(\omega)$.

\subsection{The Gounaris-Sakurai (GS) model of $F_\pi$}
The GS parametrization~\cite{Gounaris:1968mw} contains two free
parameters characterizing the $\rho$ resonance, $m_\rho$ and
$\Gamma_\rho$. Defining $k_\rho$ via $m_\rho =
2\sqrt{k_\rho^2+m_\pi^2}$, the phase shift is written
\ba\la{eq:cotd11}
&& \frac{k^3}{\omega} \cot\delta_{11}(k) = 
k^2h(\omega) - k_\rho^2 h(m_\rho) + b(k^2-k_\rho^2),
\qquad 
\\
\la{eq:b}
&& b = -\frac{2}{m_\rho}\left[\frac{2k_\rho^3}{m_\rho\Gamma_\rho}+\frac{1}{2} m_\rho h(m_\rho)
+ k_\rho^2 h'(m_\rho)\right],
\\
&& \la{eq:h_def} h(\omega) = \frac{2}{\pi} \frac{k}{\omega} \log \frac{\omega+2k}{2m_\pi}.
\ea
The form factor is then given by 
\ba
F_\pi(\omega) &=& \frac{f_0}{\frac{k^3}{\omega} \left(\cot[\delta_{11}(k)] - i\right)},
\\
f_0 &=&  -\frac{m_\pi^2}{\pi} - k_\rho^2 h(m_\rho) - b \frac{m_\rho^2}{4}. 
\ea
By analytic continuation, $F_\pi(\omega)$ is guaranteed to be unity at the origin.
In all numerical applications, we have set $m_\pi=139.57{\rm MeV}$, 
$m_\rho=773{\rm Mev}$ and $\Gamma_\rho = 130{\rm MeV}$. These values were chosen 
so as to approximately match the 2010 KLOE data~\cite{Ambrosino:2010bv}. We have 
not tried to correct for isospin breaking effects in the experimental data.

\subsection{Non-interacting pions\la{sec:freep}}
We consider here the case of non-interacting pions of mass $m_\pi$. 
The isovector current takes the form 
\be
j^a_\mu(x) = \epsilon^{abc} \pi^b(x)\partial_\mu\pi^c(x)
\ee
for pion fields with a canonically normalized kinetic term.
Then one finds in finite volume, with $E_k^2=\vec k^2+m_\pi^2$,
\be
G(x_0,L) = \frac{1}{L^3} \sum_{\vec k} k_z^2\, \frac{e^{-2E_k |x_0| }}{E_k^2}.
\la{eq:Gfv}
\ee
To evaluate the finite-volume correlator at large times, 
\eq(\ref{eq:Gfv}) is an adequate representation.
At small times however, it is more efficient to use a different representation obtained
using the Poisson formula,
\ba
G(x_0,L) &=& \frac{m_\pi^3}{6\pi^2} \sum_{\vec n}
\int_0^\infty d x\, \frac{x^4}{x^2+1} 
\\ &&
\qquad \frac{\sin (m_\pi L|\vec n| x)}{m_\pi L|\vec n| x}\, 
e^{-2m_\pi |x_0|\sqrt{x^2+1}}.
\nonumber
\ea
The $\vec n=0$ term coincides with the infinite-volume result, which
as a consistency check can also be obtained using
 Eqs.\ (\ref{eq:Gspecrep}) and (\ref{eq:RFpi}) by setting $F_\pi(\omega)=1$.
In a saddle point approximation, we have
\ba
&& G(x_0,L) -  G(x_0)\simeq   \sqrt{\frac{m_\pi}{\pi^3 |x_0|^5}} \frac{e^{-2m_\pi|x_0|}}{48}
\qquad \\ &&\nonumber
\sum_{\vec n\neq 0}  \left(3-\frac{m_\pi L^2\vec n^2}{2|x_0|}\right)\, 
\exp\left(-\frac{m_\pi L^2 \vec n^2}{4|x_0|}\right).
\ea
In this form it is clear that for fixed $L$, the expansion converges
rapidly as long as $|x_0|$ is substantially smaller than $m_\pi L^2$.
Conversely, the finite size effect is exponential for any fixed $x_0$,
but only once $L$ is a multiple of $\sqrt{|x_0|/m_\pi}$. Numerically, 
if we require that the absolute value of the exponent in the last 
exponential be at least 4, we get $x_0^{\rm max} [{\rm fm}]\simeq \frac{197}{m_\pi[{\rm MeV}]} \left(\frac{m_\pi L}{4}\right)^2$.
We also note 
that the finite-size effect is negative for large $L$.


\bibliography{/Users/harvey/BIBLIO/viscobib.bib}

\begin{thebibliography}{40}
\expandafter\ifx\csname natexlab\endcsname\relax\def\natexlab#1{#1}\fi
\expandafter\ifx\csname bibnamefont\endcsname\relax
  \def\bibnamefont#1{#1}\fi
\expandafter\ifx\csname bibfnamefont\endcsname\relax
  \def\bibfnamefont#1{#1}\fi
\expandafter\ifx\csname citenamefont\endcsname\relax
  \def\citenamefont#1{#1}\fi
\expandafter\ifx\csname url\endcsname\relax
  \def\url#1{\texttt{#1}}\fi
\expandafter\ifx\csname urlprefix\endcsname\relax\def\urlprefix{URL }\fi
\providecommand{\bibinfo}[2]{#2}
\providecommand{\eprint}[2][]{\url{#2}}

\bibitem[{\citenamefont{Blum}(2003)}]{Blum:2002ii}
\bibinfo{author}{\bibfnamefont{T.}~\bibnamefont{Blum}},
  \bibinfo{journal}{Phys.Rev.Lett.} \textbf{\bibinfo{volume}{91}},
  \bibinfo{pages}{052001} (\bibinfo{year}{2003}), \eprint{hep-lat/0212018}.

\bibitem[{\citenamefont{Gockeler et~al.}(2004)}]{Gockeler:2003cw}
\bibinfo{author}{\bibfnamefont{M.}~\bibnamefont{G\"ockeler}} \bibnamefont{et~al.}
  (\bibinfo{collaboration}{QCDSF Collaboration}), \bibinfo{journal}{Nucl.Phys.}
  \textbf{\bibinfo{volume}{B688}}, \bibinfo{pages}{135} (\bibinfo{year}{2004}),
  \eprint{hep-lat/0312032}.

\bibitem[{\citenamefont{Aubin and Blum}(2007)}]{Aubin:2006xv}
\bibinfo{author}{\bibfnamefont{C.}~\bibnamefont{Aubin}} \bibnamefont{and}
  \bibinfo{author}{\bibfnamefont{T.}~\bibnamefont{Blum}},
  \bibinfo{journal}{Phys.Rev.} \textbf{\bibinfo{volume}{D75}},
  \bibinfo{pages}{114502} (\bibinfo{year}{2007}), \eprint{hep-lat/0608011}.

\bibitem[{\citenamefont{Feng et~al.}(2011)\citenamefont{Feng, Jansen,
  Petschlies, and Renner}}]{Feng:2011zk}
\bibinfo{author}{\bibfnamefont{X.}~\bibnamefont{Feng}},
  \bibinfo{author}{\bibfnamefont{K.}~\bibnamefont{Jansen}},
  \bibinfo{author}{\bibfnamefont{M.}~\bibnamefont{Petschlies}},
  \bibnamefont{and} \bibinfo{author}{\bibfnamefont{D.~B.}
  \bibnamefont{Renner}}, \bibinfo{journal}{Phys.Rev.Lett.}
  \textbf{\bibinfo{volume}{107}}, \bibinfo{pages}{081802}
  (\bibinfo{year}{2011}), \eprint{1103.4818}.

\bibitem[{\citenamefont{Boyle et~al.}(2012)\citenamefont{Boyle, Del~Debbio,
  Kerrane, and Zanotti}}]{Boyle:2011hu}
\bibinfo{author}{\bibfnamefont{P.}~\bibnamefont{Boyle}},
  \bibinfo{author}{\bibfnamefont{L.}~\bibnamefont{Del~Debbio}},
  \bibinfo{author}{\bibfnamefont{E.}~\bibnamefont{Kerrane}}, \bibnamefont{and}
  \bibinfo{author}{\bibfnamefont{J.}~\bibnamefont{Zanotti}},
  \bibinfo{journal}{Phys.Rev.} \textbf{\bibinfo{volume}{D85}},
  \bibinfo{pages}{074504} (\bibinfo{year}{2012}), \eprint{1107.1497}.

\bibitem[{\citenamefont{Della~Morte
  et~al.}(2012{\natexlab{a}})\citenamefont{Della~Morte, Jager, Juttner, and
  Wittig}}]{DellaMorte:2011aa}
\bibinfo{author}{\bibfnamefont{M.}~\bibnamefont{Della~Morte}},
  \bibinfo{author}{\bibfnamefont{B.}~\bibnamefont{J\"ager}},
  \bibinfo{author}{\bibfnamefont{A.}~\bibnamefont{J\"uttner}}, \bibnamefont{and}
  \bibinfo{author}{\bibfnamefont{H.}~\bibnamefont{Wittig}},
  \bibinfo{journal}{JHEP} \textbf{\bibinfo{volume}{1203}}, \bibinfo{pages}{055}
  (\bibinfo{year}{2012}{\natexlab{a}}), \eprint{1112.2894}.

\bibitem[{\citenamefont{de~Divitiis et~al.}(2012)\citenamefont{de~Divitiis,
  Petronzio, and Tantalo}}]{deDivitiis:2012vs}
\bibinfo{author}{\bibfnamefont{G.}~\bibnamefont{de~Divitiis}},
  \bibinfo{author}{\bibfnamefont{R.}~\bibnamefont{Petronzio}},
  \bibnamefont{and} \bibinfo{author}{\bibfnamefont{N.}~\bibnamefont{Tantalo}},
  \bibinfo{journal}{Phys.Lett.} \textbf{\bibinfo{volume}{B718}},
  \bibinfo{pages}{589} (\bibinfo{year}{2012}), \eprint{1208.5914}.

\bibitem[{\citenamefont{Aubin et~al.}(2012)\citenamefont{Aubin, Blum,
  Golterman, and Peris}}]{Aubin:2012me}
\bibinfo{author}{\bibfnamefont{C.}~\bibnamefont{Aubin}},
  \bibinfo{author}{\bibfnamefont{T.}~\bibnamefont{Blum}},
  \bibinfo{author}{\bibfnamefont{M.}~\bibnamefont{Golterman}},
  \bibnamefont{and} \bibinfo{author}{\bibfnamefont{S.}~\bibnamefont{Peris}},
  \bibinfo{journal}{Phys.Rev.} \textbf{\bibinfo{volume}{D86}},
  \bibinfo{pages}{054509} (\bibinfo{year}{2012}), \eprint{1205.3695}.

\bibitem[{\citenamefont{Feng et~al.}(2013)\citenamefont{Feng, Hashimoto,
  Hotzel, Jansen, Petschlies et~al.}}]{Feng:2013xsa}
\bibinfo{author}{\bibfnamefont{X.}~\bibnamefont{Feng}},
  \bibinfo{author}{\bibfnamefont{S.}~\bibnamefont{Hashimoto}},
  \bibinfo{author}{\bibfnamefont{G.}~\bibnamefont{Hotzel}},
  \bibinfo{author}{\bibfnamefont{K.}~\bibnamefont{Jansen}},
  \bibinfo{author}{\bibfnamefont{M.}~\bibnamefont{Petschlies}},
  \bibnamefont{et~al.} (\bibinfo{year}{2013}), \eprint{1305.5878}.

\bibitem[{\citenamefont{Bernecker and Meyer}(2011)}]{Bernecker:2011gh}
\bibinfo{author}{\bibfnamefont{D.}~\bibnamefont{Bernecker}} \bibnamefont{and}
  \bibinfo{author}{\bibfnamefont{H.~B.} \bibnamefont{Meyer}},
  \bibinfo{journal}{Eur.Phys.J.} \textbf{\bibinfo{volume}{A47}},
  \bibinfo{pages}{148} (\bibinfo{year}{2011}), \eprint{1107.4388}.

\bibitem[{\citenamefont{Juttner and Della~Morte}(2009)}]{Juttner:2009yb}
\bibinfo{author}{\bibfnamefont{A.}~\bibnamefont{J\"uttner}} \bibnamefont{and}
  \bibinfo{author}{\bibfnamefont{M.}~\bibnamefont{Della~Morte}},
  \bibinfo{journal}{PoS} \textbf{\bibinfo{volume}{LAT2009}},
  \bibinfo{pages}{143} (\bibinfo{year}{2009}), \eprint{0910.3755}.

\bibitem[{\citenamefont{Della~Morte and Juttner}(2010)}]{DellaMorte:2010aq}
\bibinfo{author}{\bibfnamefont{M.}~\bibnamefont{Della~Morte}} \bibnamefont{and}
  \bibinfo{author}{\bibfnamefont{A.}~\bibnamefont{J\"uttner}},
  \bibinfo{journal}{JHEP} \textbf{\bibinfo{volume}{1011}}, \bibinfo{pages}{154}
  (\bibinfo{year}{2010}), \eprint{1009.3783}.

\bibitem[{\citenamefont{Dolinsky et~al.}(1991)\citenamefont{Dolinsky,
  Druzhinin, Dubrovin, Golubev, Ivanchenko et~al.}}]{Dolinsky:1991vq}
\bibinfo{author}{\bibfnamefont{S.}~\bibnamefont{Dolinsky}},
  \bibinfo{author}{\bibfnamefont{V.}~\bibnamefont{Druzhinin}},
  \bibinfo{author}{\bibfnamefont{M.}~\bibnamefont{Dubrovin}},
  \bibinfo{author}{\bibfnamefont{V.}~\bibnamefont{Golubev}},
  \bibinfo{author}{\bibfnamefont{V.}~\bibnamefont{Ivanchenko}},
  \bibnamefont{et~al.}, \bibinfo{journal}{Phys.Rept.}
  \textbf{\bibinfo{volume}{202}}, \bibinfo{pages}{99} (\bibinfo{year}{1991}).

\bibitem[{\citenamefont{Baikov et~al.}(2012)\citenamefont{Baikov, Chetyrkin,
  Kuhn, and Rittinger}}]{Baikov:2012zm}
\bibinfo{author}{\bibfnamefont{P.}~\bibnamefont{Baikov}},
  \bibinfo{author}{\bibfnamefont{K.}~\bibnamefont{Chetyrkin}},
  \bibinfo{author}{\bibfnamefont{J.}~\bibnamefont{Kuhn}}, \bibnamefont{and}
  \bibinfo{author}{\bibfnamefont{J.}~\bibnamefont{Rittinger}},
  \bibinfo{journal}{JHEP} \textbf{\bibinfo{volume}{1207}}, \bibinfo{pages}{017}
  (\bibinfo{year}{2012}), \eprint{1206.1284}.

\bibitem[{\citenamefont{Chetyrkin et~al.}(1985)\citenamefont{Chetyrkin,
  Spiridonov, and Gorishnii}}]{Chetyrkin:1985kn}
\bibinfo{author}{\bibfnamefont{K.}~\bibnamefont{Chetyrkin}},
  \bibinfo{author}{\bibfnamefont{V.}~\bibnamefont{Spiridonov}},
  \bibnamefont{and}
  \bibinfo{author}{\bibfnamefont{S.}~\bibnamefont{Gorishnii}},
  \bibinfo{journal}{Phys.Lett.} \textbf{\bibinfo{volume}{B160}},
  \bibinfo{pages}{149} (\bibinfo{year}{1985}).

\bibitem[{\citenamefont{Mallik}(1998)}]{Mallik:1997pq}
\bibinfo{author}{\bibfnamefont{S.}~\bibnamefont{Mallik}},
  \bibinfo{journal}{Phys.Lett.} \textbf{\bibinfo{volume}{B416}},
  \bibinfo{pages}{373} (\bibinfo{year}{1998}), \eprint{hep-ph/9710556}.

\bibitem[{\citenamefont{Meyer}(2009)}]{Meyer:2009kn}
\bibinfo{author}{\bibfnamefont{H.~B.} \bibnamefont{Meyer}},
  \bibinfo{journal}{JHEP} \textbf{\bibinfo{volume}{07}}, \bibinfo{pages}{059}
  (\bibinfo{year}{2009}), \eprint{0905.1663}.

\bibitem[{\citenamefont{Brandt et~al.}(2012)\citenamefont{Brandt, Della~Morte,
  Jager, Juttner, and Wittig}}]{Brandt:2012zza}
\bibinfo{author}{\bibfnamefont{B.}~\bibnamefont{Brandt}},
  \bibinfo{author}{\bibfnamefont{M.}~\bibnamefont{Della~Morte}},
  \bibinfo{author}{\bibfnamefont{B.}~\bibnamefont{J\"ager}},
  \bibinfo{author}{\bibfnamefont{A.}~\bibnamefont{J\"uttner}}, \bibnamefont{and}
  \bibinfo{author}{\bibfnamefont{H.}~\bibnamefont{Wittig}},
  \bibinfo{journal}{Prog.Part.Nucl.Phys.} \textbf{\bibinfo{volume}{67}},
  \bibinfo{pages}{223} (\bibinfo{year}{2012}).

\bibitem[{\citenamefont{Luscher}(1986)}]{Luscher:1985dn}
\bibinfo{author}{\bibfnamefont{M.}~\bibnamefont{L\"uscher}},
  \bibinfo{journal}{Commun.Math.Phys.} \textbf{\bibinfo{volume}{104}},
  \bibinfo{pages}{177} (\bibinfo{year}{1986}).

\bibitem[{\citenamefont{Luscher}(1991{\natexlab{a}})}]{Luscher:1991cf}
\bibinfo{author}{\bibfnamefont{M.}~\bibnamefont{L\"uscher}},
  \bibinfo{journal}{Nucl. Phys.} \textbf{\bibinfo{volume}{B364}},
  \bibinfo{pages}{237} (\bibinfo{year}{1991}{\natexlab{a}}).

\bibitem[{\citenamefont{Meyer}(2011)}]{Meyer:2011um}
\bibinfo{author}{\bibfnamefont{H.~B.} \bibnamefont{Meyer}},
  \bibinfo{journal}{Phys.Rev.Lett.} \textbf{\bibinfo{volume}{107}},
  \bibinfo{pages}{072002} (\bibinfo{year}{2011}), \eprint{1105.1892}.

\bibitem[{\citenamefont{Gounaris and Sakurai}(1968)}]{Gounaris:1968mw}
\bibinfo{author}{\bibfnamefont{G.}~\bibnamefont{Gounaris}} \bibnamefont{and}
  \bibinfo{author}{\bibfnamefont{J.}~\bibnamefont{Sakurai}},
  \bibinfo{journal}{Phys.Rev.Lett.} \textbf{\bibinfo{volume}{21}},
  \bibinfo{pages}{244} (\bibinfo{year}{1968}).

\bibitem[{\citenamefont{Della~Morte
  et~al.}(2012{\natexlab{b}})\citenamefont{Della~Morte, Jager, Juttner, and
  Wittig}}]{DellaMorte:2012cf}
\bibinfo{author}{\bibfnamefont{M.}~\bibnamefont{Della~Morte}},
  \bibinfo{author}{\bibfnamefont{B.}~\bibnamefont{J\"ager}},
  \bibinfo{author}{\bibfnamefont{A.}~\bibnamefont{J\"uttner}}, \bibnamefont{and}
  \bibinfo{author}{\bibfnamefont{H.}~\bibnamefont{Wittig}},
  \bibinfo{journal}{PoS} \textbf{\bibinfo{volume}{LATTICE2012}},
  \bibinfo{pages}{175} (\bibinfo{year}{2012}{\natexlab{b}}),
  \eprint{1211.1159}.

\bibitem[{\citenamefont{Wilson}(1974)}]{Wilson:1974sk}
\bibinfo{author}{\bibfnamefont{K.~G.} \bibnamefont{Wilson}},
  \bibinfo{journal}{Phys. Rev.} \textbf{\bibinfo{volume}{D10}},
  \bibinfo{pages}{2445} (\bibinfo{year}{1974}).

\bibitem[{\citenamefont{Jansen and Sommer}(1998)}]{Jansen:1998mx}
\bibinfo{author}{\bibfnamefont{K.}~\bibnamefont{Jansen}} \bibnamefont{and}
  \bibinfo{author}{\bibfnamefont{R.}~\bibnamefont{Sommer}}
  (\bibinfo{collaboration}{ALPHA collaboration}), \bibinfo{journal}{Nucl.Phys.}
  \textbf{\bibinfo{volume}{B530}}, \bibinfo{pages}{185} (\bibinfo{year}{1998}),
  \eprint{hep-lat/9803017}.

\bibitem[{\citenamefont{Luscher}(2005)}]{Luscher:2005rx}
\bibinfo{author}{\bibfnamefont{M.}~\bibnamefont{L\"uscher}},
  \bibinfo{journal}{Comput. Phys. Commun.} \textbf{\bibinfo{volume}{165}},
  \bibinfo{pages}{199} (\bibinfo{year}{2005}), \eprint{hep-lat/0409106}.

\bibitem[{\citenamefont{Luscher}(2007)}]{Luscher:2007es}
\bibinfo{author}{\bibfnamefont{M.}~\bibnamefont{L\"uscher}},
  \bibinfo{journal}{JHEP} \textbf{\bibinfo{volume}{0712}}, \bibinfo{pages}{011}
  (\bibinfo{year}{2007}), \eprint{0710.5417}.

\bibitem[{CLS(2010{\natexlab{a}})}]{CLScode}
\bibinfo{journal}{{http://luscher.web.cern.ch/luscher/DD-HMC/index.html}}
  (\bibinfo{year}{2010}{\natexlab{a}}).

\bibitem[{CLS(2010{\natexlab{b}})}]{CLS}
\bibinfo{journal}{{https://twiki.cern.ch/twiki/bin/view/CLS/WebIntro}}
  (\bibinfo{year}{2010}{\natexlab{b}}).

\bibitem[{\citenamefont{Capitani et~al.}(2012)\citenamefont{Capitani,
  Della~Morte, von Hippel, Jager, Juttner et~al.}}]{Capitani:2012gj}
\bibinfo{author}{\bibfnamefont{S.}~\bibnamefont{Capitani}},
  \bibinfo{author}{\bibfnamefont{M.}~\bibnamefont{Della~Morte}},
  \bibinfo{author}{\bibfnamefont{G.}~\bibnamefont{von Hippel}},
  \bibinfo{author}{\bibfnamefont{B.}~\bibnamefont{J\"ager}},
  \bibinfo{author}{\bibfnamefont{A.}~\bibnamefont{J\"uttner}},
  \bibnamefont{et~al.}, \bibinfo{journal}{Phys.Rev.}
  \textbf{\bibinfo{volume}{D86}}, \bibinfo{pages}{074502}
  (\bibinfo{year}{2012}), \eprint{1205.0180}.

\bibitem[{\citenamefont{Capitani et~al.}(2011)\citenamefont{Capitani,
  Della~Morte, von Hippel, Knippschild, and Wittig}}]{Capitani:2011fg}
\bibinfo{author}{\bibfnamefont{S.}~\bibnamefont{Capitani}},
  \bibinfo{author}{\bibfnamefont{M.}~\bibnamefont{Della~Morte}},
  \bibinfo{author}{\bibfnamefont{G.}~\bibnamefont{von Hippel}},
  \bibinfo{author}{\bibfnamefont{B.}~\bibnamefont{Knippschild}},
  \bibnamefont{and} \bibinfo{author}{\bibfnamefont{H.}~\bibnamefont{Wittig}},
  \bibinfo{journal}{PoS} \textbf{\bibinfo{volume}{LATTICE2011}},
  \bibinfo{pages}{145} (\bibinfo{year}{2011}), \eprint{1110.6365}.

\bibitem[{\citenamefont{Della~Morte et~al.}(2005)\citenamefont{Della~Morte,
  Hoffmann, Knechtli, Sommer, and Wolff}}]{DellaMorte:2005rd}
\bibinfo{author}{\bibfnamefont{M.}~\bibnamefont{Della~Morte}},
  \bibinfo{author}{\bibfnamefont{R.}~\bibnamefont{Hoffmann}},
  \bibinfo{author}{\bibfnamefont{F.}~\bibnamefont{Knechtli}},
  \bibinfo{author}{\bibfnamefont{R.}~\bibnamefont{Sommer}}, \bibnamefont{and}
  \bibinfo{author}{\bibfnamefont{U.}~\bibnamefont{Wolff}},
  \bibinfo{journal}{JHEP} \textbf{\bibinfo{volume}{0507}}, \bibinfo{pages}{007}
  (\bibinfo{year}{2005}), \eprint{hep-lat/0505026}.

\bibitem[{\citenamefont{Luscher et~al.}(1996)\citenamefont{Luescher, Sint,
  Sommer, and Weisz}}]{Luscher:1996sc}
\bibinfo{author}{\bibfnamefont{M.}~\bibnamefont{L\"uscher}},
  \bibinfo{author}{\bibfnamefont{S.}~\bibnamefont{Sint}},
  \bibinfo{author}{\bibfnamefont{R.}~\bibnamefont{Sommer}}, \bibnamefont{and}
  \bibinfo{author}{\bibfnamefont{P.}~\bibnamefont{Weisz}},
  \bibinfo{journal}{Nucl. Phys.} \textbf{\bibinfo{volume}{B478}},
  \bibinfo{pages}{365} (\bibinfo{year}{1996}), \eprint{hep-lat/9605038}.

\bibitem[{\citenamefont{Sint and Weisz}(1997)}]{Sint:1997jx}
\bibinfo{author}{\bibfnamefont{S.}~\bibnamefont{Sint}} \bibnamefont{and}
  \bibinfo{author}{\bibfnamefont{P.}~\bibnamefont{Weisz}},
  \bibinfo{journal}{Nucl.Phys.} \textbf{\bibinfo{volume}{B502}},
  \bibinfo{pages}{251} (\bibinfo{year}{1997}), \eprint{hep-lat/9704001}.

\bibitem[{\citenamefont{de~Divitiis et~al.}(2004)\citenamefont{de~Divitiis,
  Petronzio, and Tantalo}}]{deDivitiis:2004kq}
\bibinfo{author}{\bibfnamefont{G.}~\bibnamefont{de~Divitiis}},
  \bibinfo{author}{\bibfnamefont{R.}~\bibnamefont{Petronzio}},
  \bibnamefont{and} \bibinfo{author}{\bibfnamefont{N.}~\bibnamefont{Tantalo}},
  \bibinfo{journal}{Phys.Lett.} \textbf{\bibinfo{volume}{B595}},
  \bibinfo{pages}{408} (\bibinfo{year}{2004}), \eprint{hep-lat/0405002}.

\bibitem[{\citenamefont{Sachrajda and Villadoro}(2005)}]{Sachrajda:2004mi}
\bibinfo{author}{\bibfnamefont{C.}~\bibnamefont{Sachrajda}} \bibnamefont{and}
  \bibinfo{author}{\bibfnamefont{G.}~\bibnamefont{Villadoro}},
  \bibinfo{journal}{Phys.Lett.} \textbf{\bibinfo{volume}{B609}},
  \bibinfo{pages}{73} (\bibinfo{year}{2005}), \eprint{hep-lat/0411033}.

\bibitem[{\citenamefont{Bedaque and Chen}(2005)}]{Bedaque:2004ax}
\bibinfo{author}{\bibfnamefont{P.~F.} \bibnamefont{Bedaque}} \bibnamefont{and}
  \bibinfo{author}{\bibfnamefont{J.-W.} \bibnamefont{Chen}},
  \bibinfo{journal}{Phys.Lett.} \textbf{\bibinfo{volume}{B616}},
  \bibinfo{pages}{208} (\bibinfo{year}{2005}), \eprint{hep-lat/0412023}.

\bibitem[{\citenamefont{Jegerlehner and Nyffeler}(2009)}]{Jegerlehner:2009ry}
\bibinfo{author}{\bibfnamefont{F.}~\bibnamefont{Jegerlehner}} \bibnamefont{and}
  \bibinfo{author}{\bibfnamefont{A.}~\bibnamefont{Nyffeler}},
  \bibinfo{journal}{Phys.Rept.} \textbf{\bibinfo{volume}{477}},
  \bibinfo{pages}{1} (\bibinfo{year}{2009}), \eprint{0902.3360}.

\bibitem[{\citenamefont{Luscher}(1991{\natexlab{b}})}]{Luscher:1990ux}
\bibinfo{author}{\bibfnamefont{M.}~\bibnamefont{L\"uscher}},
  \bibinfo{journal}{Nucl.Phys.} \textbf{\bibinfo{volume}{B354}},
  \bibinfo{pages}{531} (\bibinfo{year}{1991}{\natexlab{b}}).

\bibitem[{\citenamefont{Ambrosino et~al.}(2011)}]{Ambrosino:2010bv}
\bibinfo{author}{\bibfnamefont{F.}~\bibnamefont{Ambrosino}}
  \bibnamefont{et~al.} (\bibinfo{collaboration}{KLOE Collaboration}),
  \bibinfo{journal}{Phys.Lett.} \textbf{\bibinfo{volume}{B700}},
  \bibinfo{pages}{102} (\bibinfo{year}{2011}), \eprint{1006.5313}.

\end{thebibliography}

\end{document}